\documentclass[aps,pre,groupedaddress,twocolumn]{revtex4}

\newcommand{\Uc}{\mathcal{U}}

\usepackage{hyperref}
\usepackage{subfigure}
\usepackage{color}
\usepackage{graphicx}

\begin{document}
\expandafter\ifx\csname urlprefix\endcsname\relax\def\urlprefix{URL }\fi

\title{\large
 Droplets moving on a fluid surface: interference pattern from two slits }

\large
\author{Valeriy I. Sbitnev}
\email{valery.sbitnev@gmail.com}
\address{B. P. Konstantinov St. Petersburg Nuclear Physics Institute, NRC Kurchatov Institute, Gatchina, Leningrad district, 188350, Russia;\\
 Department of Electrical Engineering and Computer Sciences, University of California, Berkeley, Berkeley, CA 94720, USA
}


\date{\today}

\begin{abstract}
 The Feynman path integral approach for solving the motion of a droplet
 along a silicon oil surface is developed by replacing the Planck
 constant by a surrogate parameter. The latter is proportional to the
 surface tension of the silicon oil multiplied by the area of the thin
 air film, separating the droplet from the oil, and by the half-period of
 the Faraday oscillations. It is shown that the Navier-Stokes equation
 together with the mass conservation equation can be reduced to the
 Schr\"{o}dinger equation when the surrogate parameter replaces the
 Planck constant. The Feynman path integral underlying the
 Schr\"{o}dinger equation is used then to calculate a wave function that
 plays the role of the de Broglie pilot-wave. \\


\end{abstract}

\maketitle

\large

\section{\label{sec:level1}Introduction.}

\begin{figure}[htb!]
  \centering
  \begin{picture}(200,280)(30,15)
  \includegraphics{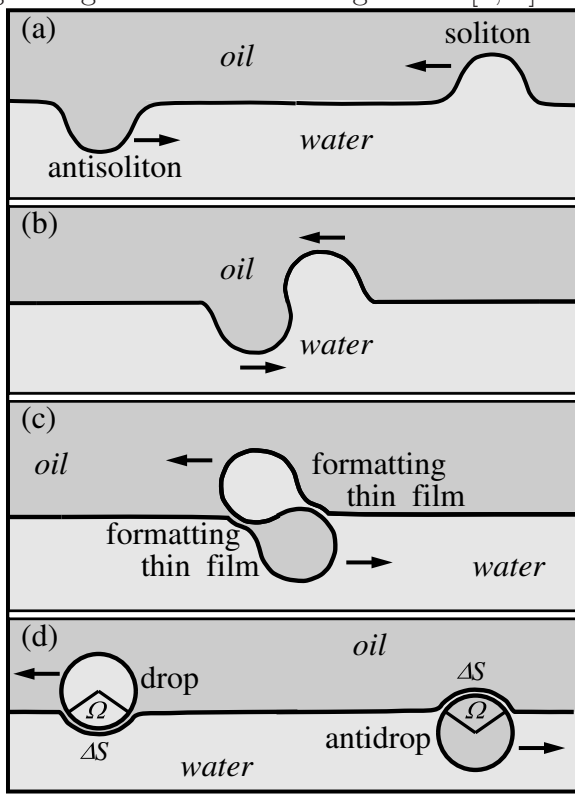}
  \end{picture}
  \caption{
  Creation of the pair "drop-anti-drop" as a result of collision of two solitary waves
 on the interface of media "oil-water"~\cite{Krenev}: (a) two solitary waves, soliton and
 antisoliton, move to meet each other; (b) moment of their collision; (c) formation of
 thin films between droplets and fluids; (d) droplet and anti-droplet are separated from
 fluids by thin films with area about~${\Delta S}$.  }
  \label{fig=1}
\end{figure}

 Recently a team of French scientists has shown that oil droplets on the
 "silicon oil - air" interface can behave as quantum particles,
 demonstrating the interference phenomenon from two slits~\cite{CouderForte:2006}. Behavior
 of the droplets leads to the emergence of amazing
 structures~\cite{CouderEtAll:2005, ProtiereEtAll:2006, EddiEtAll:2011},
 appearance of which is typical for self-organization of interacting
 atoms as quantum objects. One more quantum-mechanical analogy is the
 creation of a pair "drop - anti-drop" as a result of the collision of
 two solitary waves~\cite{Krenev}, Fig.~\ref{fig=1}, thus simulating the phenomenon of
 "electron-positron" pair creation.
 Feynman diagram technique~\cite{Feynman1998} involving operators of creation and annihilation
 of different harmonic modes~\cite{DorboloEtAl2008, GiletEtAl2008}
 can be used  in the same manner in order to observe
 the creation/annihilation of drop-anti-drop pairs.

\begin{figure}[htb!]
  \centering
  \begin{picture}(200,70)(20,25)
  \includegraphics{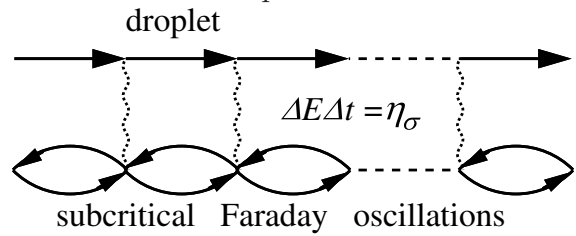}
  \end{picture}
  \caption{
  Feynman-like diagram representing the motion of a droplet bouncing
  along the silicon oil surface. Exchange by energy via a generated
  surface wave occurs at each bounce. Here relation
  ${\Delta E}{\Delta t} = \eta_{\sigma}$ simulates
  the quantum-mechanical relation
  ${\Delta E}{\Delta t} = \hbar$.
 Similar diagram technique~\cite{Feynman1998} involving operators of creation and annihilation
 of different harmonic modes~\cite{DorboloEtAl2008, GiletEtAl2008}
 can be used  in order to observe the creation/annihilation of drop-anti-drop pairs.
}
\label{fig=2}
\end{figure}

 Generating Faraday waves by vertically vibrating bath is an
 indispensable part in such
 experiments~\cite{CouderForte:2006, CouderEtAll:2005, ProtiereEtAll:2006, EddiEtAll:2011, MilesHenderson1990, Miles1993, PerinetEtAll2009}.
 Note that these waves
 are slightly below the bifurcation threshold. Such oscillations can be
 imagined to be akin to vacuum zero-point oscillations. In other words,
 the motion of droplets on such surfaces may imitate the motion of
 particles in the vacuum, Fig.~\ref{fig=2}.

 Notwithstanding the similarity with quantum-mechanical phenomena we
 cannot apply quantum formulas to describe the observed motion first of
 all because of the smallness of the main quantum constant - the Planck
 constant $h~\sim~10^{-34}$~J$\cdot$s. The formula ${\Delta E}{\Delta t} = \hbar$
 can relate to the exchange
 of energy with the virtual particles of quantum vacuum, not with the
 subcritical Faraday oscillations. Droplets are heavy objects having a
 mass about $m \sim 1$ mg and moving with velocities about 10 mm/s~\cite{CouderEtAll:2005}.
 Formally, we can evaluate that wavelength of such droplets would be
 about $10^{-23}$ mm. From here it follows that the Planck constant cannot be
 adopted as a native constant for the droplet interference experiments.
 Instead, we need to use a surrogate parameter, $\eta_{\sigma}$, replacing the Planck
 constant,~Fig.~\ref{fig=2}.

 Let us recall experiment with silicon droplets~\cite{CouderForte:2006}.
 A bouncing droplet
 rests on the fluid surface divided by a thin air film which prevents
 coalescence of the droplet with the fluid in the bath~\cite{CouderEtAll:2005} (it is akin to
 retention of water striders on a water surface due to surface tension).
 So, the surface tension of the silicon oil, $\sigma = 0.0209$~N/m~\cite{ProtiereEtAll:2006}, can be
 the main parameter determining the surrogate parameter. On the other
 hand, the droplet rests on the basic fluid substance of a small area,
 ${\Delta S}$, restricted by a solid angle $\Omega$, Fig.~\ref{fig=1}(d). The energy consumed to
 support the droplet on the fluid surface for duration of bouncing, ${\Delta t}$,
 is ${\Delta E} = \sigma{\Delta S}$. From here it follows,
 that the surrogate parameter, $\eta_{\sigma}$, can
 have the following form
\begin{equation}\label{eq=1}
    \eta_{\sigma} = {\Delta E}{\Delta t} = \sigma{\Delta S}T_{0}/2.
\end{equation}
 Here ${\Delta t} = T_{0}/2$ is a half-period of the Faraday oscillations of the fluid
 generated by an external vibrator~\cite{CouderForte:2006}. Estimation gives $\eta_{\sigma} \sim 10^{-11}$ J$\cdot$s
 for ${\Delta S}$ being represented by 0.05 part of the surface area of a sphere of
 a droplet having diameter about 0.76~mm~\cite{EddiEtAll:2011} and a period of the Faraday
 oscillations $T_{0} \sim 12$ ms (the forcing frequency is 80~Hz~\cite{EddiEtAll:2011}). Now we may
 evaluate a wavelength of the droplet, $\lambda$, having mass $m \sim 0.2$ mg and
 moving with a velocity, $v_{z}$, of about 10 mm/s. It is about $\eta_{\sigma}/mv_{z} = 5$ mm.
 This wavelength is in a good agreement with that given in~\cite{CouderForte:2006, EddiEtAll:2011}.

 By adopting the parameter $\eta_{\sigma}$ as a quantum of action we show that the
 Navier-Stokes equation together with the mass conservation equation has
 a close relation to the Schr\"{o}dinger equation, where the Planck constant
 should be replaced by this parameter. Next we calculate the Feynman path
 integral for the motion of droplets through a double-slit grating.
 Calculations disclose an interference pattern behind the grating and a
 set of Bohmian trajectories along which the motion of the droplets
 occurs.

\section{\label{sec:level2}Transformation of the Navier-Stokes Equations.}

 Quantum theory in a hydrodynamic form was formulated by Erwin Madelung
 in 1926~\cite{Madelung1926} as an alternative formulation of the Schr\"{o}dinger equation.
 Remarkably that Madelung's equations exhibit a close relationship
 through the Bohmian mechanics~\cite{GuantesEtAll2004, OriolsEtAll2012}
 with hydrodynamic equations such
 as the Navier-Stokes equations. It gives the reason to hypothesize that
 quantum medium behaves like a fluid with irregular fluctuations~\cite{BohmVigier1952}.
 On the other hand, we may suppose that behavior of an incompressible liquid
 can be described by a Schr\"{o}dinger-like equation with a special parameter
 replacing the Planck constant. Let us trace this supposition.

 Two equations, describing flow of an incompressible fluid, are~\cite{LandauLifshitz1987}:\\
 (a) the Navier-Stokes equation
\begin{equation}\label{eq=2}
  \rho_{M}
  \Biggl(
    {{\partial\,{\vec v}}\over{\partial\,t}}
     +({\vec v}\cdot\nabla){\vec v}
  \Biggr)
  =  -{\nabla P} + \mu{\nabla^{2}{\vec v}} + {{\vec F}\over{\Delta V}},
\end{equation}
 (b) the mass conservation equation
\begin{equation}\label{eq=3}
    {{\partial\,{\rho_{M}}}\over{\partial\,t}}
     + \nabla(\rho_{M}{\vec v})    = 0.
\end{equation}
 Here ${\vec v}$ is a flow velocity, $P$ is a pressure, $\mu$ is the viscosity
 coefficient, and ${\vec F}/{\Delta V}$ is a force per unit volume ${\Delta V}$.
 A mass density $\rho_{M}$ is
 defined as the mass of the fluid, $M$, per volume ${\Delta V}$:
\begin{equation}\label{eq=4}
  \rho_{M} = {{M}\over{{\Delta V}}} = {{mN}\over{{\Delta V}}} = m\rho.
\end{equation}
 Next we shall represent $M$ as a product of an elementary mass m by the
 number of these masses, $N$, contained within volume ${\Delta V}$. Then the mass
 density $\rho_{M}$ can be defined as a product of the elementary mass $m$ by the
 density of elementary carriers, $\rho=N/{\Delta V}$. The elementary carrier is a
 droplet-like inhomogeneity, which moves with the local stream velocity
 of the equivalent fluid. We assume that each elementary carrier of mass
 $m$ is subjected to the Brownian motion with a diffusion coefficient
 inversely proportional to $m$ and no friction~\cite{Nelson1966}. In this sense, $\rho$
 represents a probability density of finding the carrier within the
 volume ${\Delta V}$ and obeys the conservation law~(\ref{eq=3}).

 As for the term $({\vec v}\cdot\nabla){\vec v}$ in Eq.~(\ref{eq=2}),
 it can be rewritten as
 $({\vec v}\cdot\nabla){\vec v}={\nabla v^{2}/2}-[{\vec v}\times[\nabla\times{\vec v}]]$.
 Further we shall
 consider irrotational flows, that is, $[\nabla\times{\vec v}]]=0$.
 Let the force be conservative, ${\vec F} = -\nabla \Uc$. In this
 case we rewrite the Navier-Stokes equation in the following form
\begin{equation}\label{eq=5}
    m {{\partial\,{\vec v}}\over{\partial\,t}}
  + m \nabla {{v^{2}}\over{2}}
  - {{{\nabla P}}\over{\rho}}
    + {{\mu}\over{\rho}}\nabla^{2} {\vec v}
    - {{\nabla \Uc}\over{\rho\Delta V}}.
\end{equation}
 The rightmost term, $-\nabla\Uc/\rho\Delta V = -\nabla U$,
 is the force acting on the elementary carrier.

 One can see that $m{\vec v}$ is a momentum of the elementary carrier and $mv^{2}/2$
 is its kinetic energy. Let us define these quantities through
 introducing action $S$ - a mathematical functional which accounts for the
 history of the system in the Lagrangian
 mechanics~\cite{GuantesEtAll2004, OriolsEtAll2012}:
\begin{eqnarray}\label{eq=6}
  {\vec p} &=& m{\vec v} = \nabla S,
\\
  m{{v^{2}}\over{2}} &=& {{1}\over{2m}}(\nabla S)^{2}.
\label{eq=7}
\end{eqnarray}
 Substituting these expressions
 in the Navier-Stokes equation~(\ref{eq=5}) we obtain
\begin{eqnarray} \nonumber
 &&
    \nabla
    \Biggl(
    {{\partial\,S}\over{\partial\,t}}
  + {{1}\over{2m}}(\nabla S)^{2}
  + U + Q
     \Biggr)
\\ &&
 =
  - {{{\nabla P}}\over{\rho}}
    + {{\mu}\over{\rho}}\nabla^{2} {\vec v}
    + {\nabla Q}.
\label{eq=8}
\end{eqnarray}
 We have added gradient of the term
\begin{equation}\label{eq=9}
  Q =
  -{{\eta_{\sigma}^{2}}\over{2m}}
  \Biggl[
   {{\nabla^{2}\rho}\over{2\rho}}
   -\biggr(
     {{\nabla\rho}\over{2\rho}}
    \biggl)^{2}\;
  \Biggr]
\end{equation}
 to both sides of Eq.~(\ref{eq=8}).
 If $\eta_{\sigma}$ will be substituted by $\hbar$, the term $Q$ will represent
 the quantum potential first defined by D. Bohm~\cite{Bohm1952}.

 Right hand side (RHS) of Eq.~(\ref{eq=8}) besides the pressure force and the viscosity force
 contains also the term~(\ref{eq=9}). Under the assumption of incompressibility,
 the density of the elementary fluid volume is constant. It stays below
 the Faraday instability threshold. From here we can assume that the
 gradient of $\rho$ is zero and we may omit the gradient operator in
 Eq.~(\ref{eq=8}).
 RHS, free from the gradient, looks as
\begin{equation}\label{eq=10}
  - {{P}\over{\rho}}
    + {{\mu}\over{\rho}}\nabla {\vec v}
    -{{\eta_{\sigma}^{2}}\over{2m}}
  \Biggl[
   {{\nabla^{2}\rho}\over{2\rho}}
   -\biggr(
     {{\nabla\rho}\over{2\rho}}
    \biggl)^{2}
  \Biggr]
  = -V_{N}.
\end{equation}
 The Fick's law says that the diffusion flux, $J$, is proportional to the
 negative value of the density gradient, $J=-(D/2)\nabla\rho$ , where
 $D = \eta_{\sigma}/2m$ is the diffusion coefficient~\cite{Nelson1966}.
 The term $\eta_{\sigma}\nabla J$ has dimensions of the pressure.
 From here it follows that the pressure $P$ has a diffusion nature
\begin{equation}\label{eq=11}
  P = \eta_{\sigma}\nabla J = - {{\eta_{\sigma}^{2}}\over{4m}}\nabla^{2}\rho.
\end{equation}
 By substituting this expression in Eq.~(\ref{eq=10}) we have
\begin{equation}\label{eq=12}
  V_{N} = -{{\eta_{\sigma}^{2}}\over{2m}}
        \Biggl(
          {{\nabla\rho}\over{2\rho}}
        \Biggr)^{2}
        - {{\mu}\over{m\rho}}\nabla^{2} S.
\end{equation}
 By integrating Eq.~(\ref{eq=8}) over the volume of the fluid we get
\begin{eqnarray} \nonumber
 && {{\partial\,S}\over{\partial\,t}}
 + {{1}\over{2m}}(\nabla S)^{2} + U + V_{N}
\\ 
 && - {{\eta_{\sigma}^{2}}\over{2m}}
      \Biggl(
         {{\nabla^{2}\rho}\over{2\rho}}
      \Biggr)
    +{{\eta_{\sigma}^{2}}\over{2m}}
      \Biggl(
         {{\nabla\rho}\over{2\rho}}
      \Biggr)^{2} = C,
\label{eq=13}
\\
 && {{\partial\,\rho}\over{\partial\,t}}
  + \nabla(\rho\cdot{\vec v}) = 0.
\label{eq=14}
\end{eqnarray}
 Here we wrote down also the continuity equation for the probability density $\rho$.
 In Eq.~(\ref{eq=13}) $C$ is an integration constant.

\begin{figure*}[htb!]
  \centering
  \begin{picture}(200,380)(130,10)
  \includegraphics{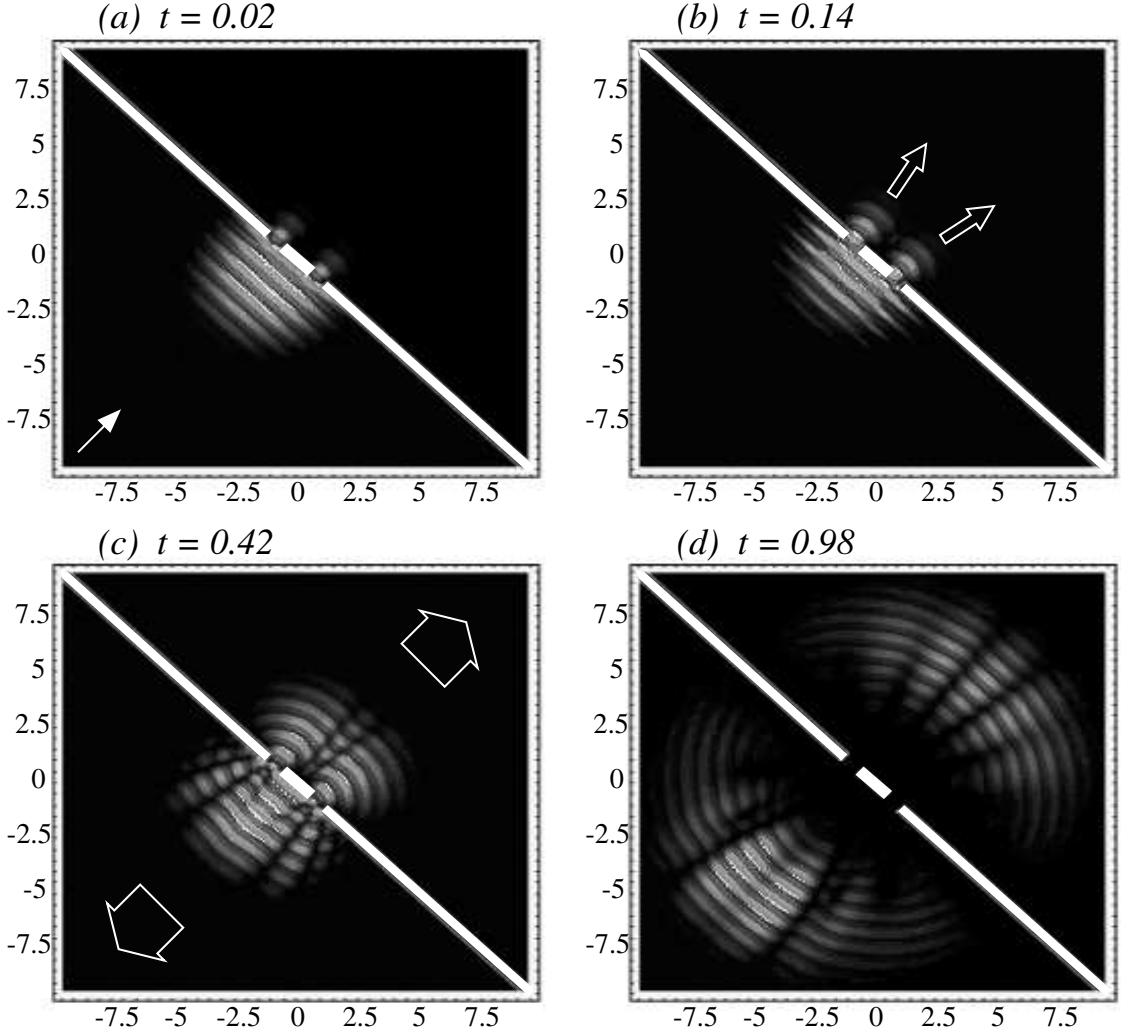}
  \end{picture}
  \caption{
 Scattering of a Gaussian soliton-like wavepacket on a barrier
 containing two slits~\cite{Sbitnev2009}.
 Time here is in seconds and scale of the area is given in centimeters.
 Velocity of the wave is about 150 mm/s. It corresponds to the phase
 velocity of the Faraday waves~\cite{CouderForte:2006}.}
  \label{fig=3}
\end{figure*}

 The modified Hamilton-Jacobi equation~(\ref{eq=13}) with extra terms
 and the continuity equation~(\ref{eq=14}) stems from a nonlinear Schr\"{o}dinger like equation
\begin{equation}\label{eq=15}
  {\bf i} {\eta_{\sigma}}
  {{\partial\,\Psi}\over{\partial\,t}}
 = -{{\eta_{\sigma}^{2}}\over{2m}} {\Delta\Psi}
 + U\Psi + V_{N}\Psi - C\Psi.
\end{equation}
 Nonlinearity arises due to the term $V_{N}$ that depends on both $\rho$ and $S$.
 Note that the wave function $\Psi$ written down in a polar form contains exactly
 these functions
\begin{equation}\label{eq=16}
  \Psi = \sqrt{\rho}\exp\{{\bf i}S/{\eta_{\sigma}}\}.
\end{equation}
 By substituting this function in Eq.~(\ref{eq=15}) and by separating real and imaginary parts,
 we obtain Eqs.~(\ref{eq=13}) and~(\ref{eq=14}).

 Due to the assumption that the fluid is incompressible and below the
 Faraday instability threshold, the term $V_{N}$ in Eq.~(\ref{eq=13}) can
 be omitted. We come to the linear Schr\"{o}dinger equation. A solution
 of the linear Schr\"{o}dinger equation with the potential simulating a
 barrier with two slits~\cite{Sbitnev2009} is shown in Fig.~\ref{fig=3}.
 Scattering of a Gaussian soliton-like wave on the slits produces two
 waves - reflected and transmitted. Both waves evolve against the
 background of subcritical Faraday waves.
 Similar interference  manifestation of waves on the water surface
 can be found on YouTube.
 Here is one~\cite{WaterWaveInterference}.

 As for the Faraday waves,
 Eq.~(\ref{eq=15}) under conditions stated
 in~\cite{MilesHenderson1990, Miles1993}
 reduces to the Miles-Henderson wave equation describing theirs.

\section{\label{sec:level3}Moving droplet in terms of the path integral.}

 Note that the Schr\"{o}dinger equation can be deduced from the Feynman
 path integral~\cite{Feynman1948} by successive expansion in a Taylor
 series of kernels of this integral~\cite{FeynmanHibbs1965, Derbes1995}. It means
 that along with the Schr\"{o}dinger equation the path integral can be
 used for consideration of a moving droplet through
 slits~\cite{CouderForte:2006}.
\begin{figure}[htb!]
  \centering
  \begin{picture}(200,140)(20,15)
  \includegraphics{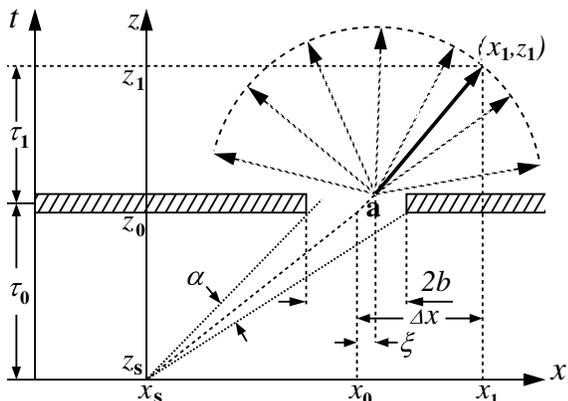}
  \end{picture}
  \caption{
  Single slit is localized in the vicinity of point $x_0$ on a grating
  situated at distance $z_0-z_s$ away from a source. A droplet, emitted
  by the source in the vicinity of point $(x_s, z_s)$, moves along a path
  within a ray $\alpha$ to the slit and further along some path pointed
  to by arrows.}
\label{fig=4}
\end{figure}
 A main step here is the replacement of
 the Planck constant $\hbar$ in the path integral by the surrogate
 parameter $\eta_{\sigma}$ equal to $1.185\times10^{-11}$ J$\cdot$s in
 our case. Its value is evaluated for a case of the forcing frequency
 $f_{0}$ = 80~Hz and the droplet diameter $D = 0.76$~mm~\cite{EddiEtAll:2011}.
 From here it follows that at $\rho_{M} = 965$~kg/m$^3$ the mass of the droplet is
 $m = \rho_{M} (4/3)\pi(D/2)^3 \sim 0.22$~mg. Let us begin from presentation
 of a Lagrange function that describes the free motion of the droplet from
 a source localized at $(x_s, z_s)$ through $(x_1, z_1)$ within a slit and
 further to a zone of detection, $(x_2, z_2)$, as shown in
 Fig.~\ref{fig=4}.
 There is no any interaction with other droplets.
 The Lagrange function of the free motion is
\begin{equation}\label{eq=17}
  L = m{{{\dot x}^{\,2}}\over{2}} + {\rm const}.
\end{equation}
 Here $m$ is the mass of the droplet and ${\dot x}$ is its transversal
 velocity. Its longitudinal velocity, $v_{z}$, can take values from 1~mm/s up
 to 20~mm/s~\cite{CouderForte:2006}. By moving the droplet by a small
 distance ${\delta x} = (x_{b}-x_{a}) \ll 1$ in the transversal
 direction, performed for a small time increment
 ${\delta t} = (t_{b}-t_{a}) \ll 1$, we
 write down a weight factor
 $\exp\{ {\bf i}S/{\eta_{\sigma}} \} = \exp\{ {\bf i}L{\delta t} /{\eta_{\sigma}} \}$
 of the path integral in the following form~\cite{FeynmanHibbs1965}
\begin{equation}\label{eq=18}
 {\rm e}^{\,{\bf i}L{\delta t}/{\eta_{\sigma}}}
 = \exp
   \Biggl\{
    {{{\bf i}m(x_{b}-x_{a})^{2}}\over{2\eta_{\sigma}(t_{b}-t_{a})}}
   \Biggr\}
\end{equation}
 The path integral computes probability amplitude that is equal to integral convolution
 of two kernels, each describing the motion of the free particle (the first describes
 the motion from the source to the slit, and the second describes the motion behind the
 slit, Fig.~\ref{fig=4}:
\begin{widetext}
\begin{equation} \label{eq=19}
  \psi(x_1,x_0,x_s) =
  \int\limits_{-b}^{b}
    K(x_1,\tau_0+\tau_1;x_0+\xi,\tau_0)
    K(x_0+\xi,\tau_0;x_s,0)d\xi.
\end{equation}
 The kernel consists of the weight factor~(\ref{eq=18})
 multiplied by an amplitude factor~\cite{FeynmanHibbs1965}
\begin{equation}\label{eq=20}
  K(x_{b},t_{b};x_{a},t_{a}) =
  \Biggl[
    {{2\pi{\bf i}\eta_{\sigma}(t_{b}-t_{a})}\over{m}}
  \Biggr]^{-1/2}
  \cdot \exp
  \Biggl\{
   {{{\bf i}m(x_{b}-x_{a})^2}\over{2\eta_{\sigma}(t_{b}-t_{a})}}
  \Biggr\}.
\end{equation}
 After all computations of the path integral in approximation of the slit by
 a single Gaussian curve~\cite{Sbitnev2010}, we find a wave function from a single slit
\begin{eqnarray} \nonumber
 && \psi(x_1,z_1,x_0,z_0,x_s,z_s) =
    \sqrt{ {{m}\over{2\pi{\bf i}\eta_{\sigma}\tau_0\tau_1}} }
    \Biggl(
      {{1}\over{\tau_1}}
     +{{1}\over{\tau_0}}
     +{\bf i} {{\eta_{\sigma}}\over{m b^{2}}}
    \Biggr)^{-1/2}
\\ \label{eq=21}
 && \times  \exp
    \Biggl\{
     {{{\bf i}m}\over{2\eta_{\sigma}}}
    \Biggl(
       {{(x_1-x_0)^2}\over{\tau_1}}
     + {{(x_0-x_0)^s}\over{\tau_0}}
     - {{(
         (x_1-x_0)/\tau_1-(x_01-x_s)/\tau_0
         )^2}\over{(1/\tau_1+1/\tau_0 + {\bf i}\eta_{\sigma}/m b^2)}}
    \Biggr)
    \Biggr\}
\end{eqnarray}
 Here the time values, $\tau_0$ and $\tau_1$, relate to the coordinates
 $z_s$, $z_0$, $z_1$ according to the following formulas:
 $\tau_0 = (z_0-z_s)/v_z$ and $\tau_1 = (z_1-z_0)/v_z$, see Fig.~\ref{fig=4}.
 Here $v_z$ is the longitudinal velocity.
\end{widetext}

 For the sake of simplicity we remove the source to negative infinity.
 In this case $\tau_0$ tends to infinity and the amplitude factor
 $A= (m/2\pi{\bf i}\eta_{\sigma}\tau_0\tau_1)^{1/2}$ tends to zero
 (luminosity of a remote source tends to zero).
 In the paraxial approximation (sources of the droplets are removed to infinity,
 but $A$ remains finite) the wave function looks
 as~\cite{Sbitnev2011, Sbitnev2013}
\begin{eqnarray} \nonumber
&&    \psi(x,z,x_0,z_0) =
\\
&&    A{\mit\Sigma}_{S}^{-1/2}
    \exp
    \Biggl\{
      {\bf i}\pi{{(x-x_0)^2}\over{\lambda(z-z_0)}}
    \Biggl(
      1 - {{1}\over{\mit\Sigma_{S}}}
    \Biggr)
    \Biggr\}.
\label{eq=22}
\end{eqnarray}
 Here $A$ is the amplitude factor and the term
\begin{equation}\label{eq=23}
   \mit\Sigma_{S} = 1 +
    {\bf i}{{\lambda(z-z_0)}\over{2\pi b^{2}}}
\end{equation}
 is a dimensionless complex-valued distance-dependent spreading which
 relates closely to the complex-valued time-dependent spreading~\cite{SanzMiret-Artes2007, SanzMiret-Artes2008}.
 Note that instead of parameters $m$, $v_z$, and $\eta_{\sigma}$
 a wavelength $\lambda = \eta_{\sigma}/mv_z$
 appears in Eqs.~(\ref{eq=22})-(\ref{eq=23}).
 Also in Eq.~(\ref{eq=23}) $b$ is half-width of the slit.
 Hereinafter we omit subscript 1: $(x_1, z_1) \rightarrow (x, z)$. Therefore $(x_0, z_0)$
 and $(x, z)$ describe the location of the center of the grating and a
 position of the detecting droplet, respectively. Now we can write down a
 general wave function from the grating as superposition of the wave
 functions~(\ref{eq=22}) from $N$ slits:
\begin{equation}\label{eq=24}
 |\Psi(x,z)\rangle = {{1}\over{N}}
 \sum\limits_{k=-{{N-1}\over{2}}}^{k={{N-1}\over{2}}}
 \psi(x-kd,z).
\end{equation}
 Here $d$ is the distance between the slits.
 In case of even number of the slits, $N = 2K$,
 $k$ runs $-K+1/2, -K+3/2, \ldots, -1/2, 1/2, \ldots , K-3/2, K-1/2$.
 For odd number, $N = 2K+1$,
 $k$ runs $-K, -K+1, \ldots, -1, 0, 1, \ldots, K-1, K$.
 We have chosen here a reference frame with the origin placed in $x_0 = 0$ and $z_0 = 0$.

 Probability density function in the vicinity of the observation point $(x, z)$ reads
\begin{equation}\label{eq=25}
    p(x,z) = \langle \Psi(x,z) | \Psi(x,z) \rangle.
\end{equation}
 In order to evaluate these computations we shall consider scattering of
 droplets on a grating containing two slits, $N = 2$, with $d = 6b$ and the
 width of the slits $2b = 10$~ mm.
\begin{figure}[htb!]
  \centering
  \begin{picture}(200,110)(10,15)
  \includegraphics{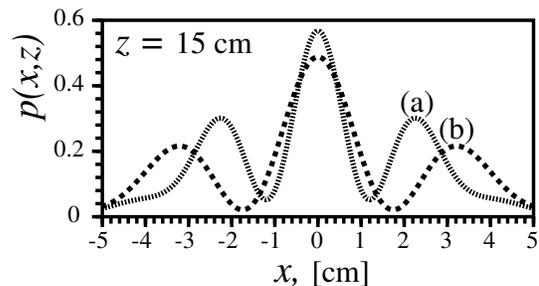}
  \end{picture}
  \caption{
 Interference fringes in a cross-section of the probability density $p(x,z)$
 in the far-field, $z = 15$~cm, : (a)~$\lambda = 4.75$~mm; (b)~$\lambda = 6.95$~mm.}
\label{fig=5}
\end{figure}
 These parameters will be fixed for all
 wavelengths in order to compare any output data. Two wavelengths,
 $\lambda = 4.75$~mm and $\lambda = 6.95$~mm~\cite{CouderForte:2006, EddiEtAll:2011},
 are chosen for subsequent
 consideration. Cross-sections of the probability density function
 disclose interference fringes in the far-field, Fig.~\ref{fig=5}, arising after
 passing droplets through an obstacle with two slits. The two curves, (a)
 for $\lambda = 4.75$~mm and (b) for $\lambda = 6.95$~mm, show qualitative accordance
 with those presented in~\cite{CouderForte:2006}. One can see that as the wavelength
 increases the fringes diverge apart. There is, however, some
 discrepancy. It is due to the fact that the formula in~\cite{CouderForte:2006}
 describes amplitude of the diffraction, whereas Eq.~(\ref{eq=25}) relates to
 description of the intensity. In fact, we need to square an observed amplitude
 in order to get a clear description of the inter ference pattern
 in the far-field~\cite{Sbitnev2011}.

 As follows from observations~\cite{EddiEtAll:2011} a droplet induces a wave on surface
 of the oil at the time of each rebound. The wave retains memory about previous impacts
 of the droplet and corrects its subsequent motion. It is a manifestation of the effect
 of the de Broglie pilot-wave that guides the droplet along an optimal path. In our case
 the wave function $|\Psi(x,z)\rangle$ represents the de Broglie pilot-wave, and the
 optimal path is called the Bohmian trajectory~\cite{GuantesEtAll2004, OriolsEtAll2012}.
 Velocity of a shift of the droplet in the plane $(x, z)$
 is computed from Eq.~(\ref{eq=6})
\begin{equation}\label{eq=26}
 {\dot{\vec r}} = {{1}\over{m}}\nabla S
    = {{\eta_{\sigma}}\over{m}}\Im[\Psi^{-1}\nabla\Psi].
\end{equation}
 The wave function represented in a polar form,
 $\Psi = R\cdot\exp\{{\bf i}S/\eta_{\sigma}\}$,
 verifies this formula.
\begin{figure}[htb!]
  \centering
  \begin{picture}(200,240)(20,10)
  \includegraphics{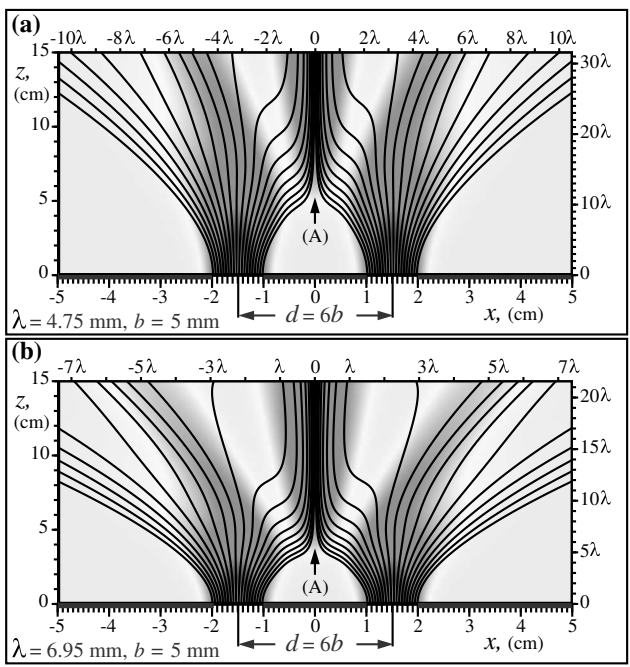}
  \end{picture}
  \caption{
   Bohmian trajectories going out from two slits are shown by black curves against
 the background of the probability density depicted by gray (light gray refers to low
 density; dark gray refers to high density):
 (a)~$\lambda = 4.75$~mm;
 (b)~$\lambda = 6.95$~mm.
 Arrows (A) point to places where Bohmian trajectories dramatically change direction. }
  \label{fig=6}
\end{figure}
 Here $R$ is its amplitude and $S$ divided by $\eta_{\sigma}$
 is a phase. Knowing the velocity from Eq.~(\ref{eq=26}) we can find a mean value of
 the optimal path~\cite{Sbitnev2013} - the Bohmian-like trajectory. The trajectories
 are shown in Figs.~\ref{fig=6}(a) and~\ref{fig=6}(b) by black curves against the background
 of the probability density colored in gray (it ranges from light gray
 (low density) to dark gray (high density)). Fig.~\ref{fig=6}(a) shows a bundle of
 the trajectories for the case of $\lambda = 4.75$~mm; the current velocity,
 $\eta_{\sigma}/(\lambda m)$, is about 11~mm/s.
 Fig.~\ref{fig=6}(b) shows a bundle of the trajectories
 for the case of $\lambda = 6.95$~mm; the current velocity is about 8~mm/s.
 One can see that at the increasing wavelength the trajectories diverge apart
 considerably stronger for the same path length along $z$.
 One can see, in particular, that for observation of the interference effects,
 the size of the bath can be about $10\times15$~cm$^2$.
 Some attention should be drawn to the area pointed to by arrows (A) in
 Fig.~\ref{fig=6}. Here the Bohmian-like trajectories change directions
 dramatically. A possible mechanism of such a deviation of the
 trajectories can be as follows~\cite{CouderForte:2006, EddiEtAll:2011}:
 (i)~the droplet moving through a slit induces a Faraday wave;
 (ii)~velocity of the wave is an order of magnitude larger than the velocity of the droplet;
 (iii)~for that reason the Faraday wave has time to reach the second slit;
 (iv)~it induces a secondary Faraday wave from this slit;
 (v)~the latter has time to reach the area pointed by arrow (A).
 Because of this the wave brings a correction in the motion of the droplet in the
 vicinity of this place. Both primary and secondary waves, in the superposition,
 play a role of the guiding wave. It should be noted that there are no intersections
 of the Bohmian trajectories with each other as they go on. Intersections of
 traces shown in a figure in~\cite{CouderForte:2006} are in contradiction with the
 ideas of de Broglie and Bohm~\cite{Bohm1952, DeBrogile1953, BohmHiley1982}. 

\section{\label{sec:level4}Conclusion.}

 Droplets bouncing on a vibrating silicon oil surface behave themselves as particles, if the vibrations are supported slightly below the Faraday instability threshold. Such vibrations are akin to zero-point vacuum oscillations. In this case we may use the Feynman path integral to study interference effects induced by moving droplets through obstacles having slits. However, instead of the Planck constant we need to use a surrogate parameter which is material dependent and also depends on the forcing oscillations and the size of the droplets.

 In the vicinity of the Faraday instability threshold the oil surface is very sensitive to weak contacts. Owing to nearness to the Faraday instability threshold, excited waves are long-lived. It is manifestation of a memory effect on the moving bouncing droplet~\cite{EddiEtAll:2011}. Superposition of such excited waves makes the de Broglie pilot wave~\cite{OriolsEtAll2012} guiding the droplet along an optimal path.

 Basic foundation for description of the de Broglie pilot wave goes back to the equations of Navier-Stokes and the conservation of mass. In approximation of the irrotational incompressible fluid these two equations are reduced to the nonlinear Schr{\"o}dinger equation, wherein the surrogate parameter is used instead of the Planck constant. Its wave solutions are determined by boundary conditions imposed on a problem under consideration, among which a grating with slits plays the crucial role. The latter gives a Fourier image of the grating in the far-field, which forms a virtual interference pattern from the slits. In turn, the moving droplet generates circular waves at each impact with the fluid surface. These waves, having a greater velocity of spreading, bring the interference pattern along the motion of the droplet. It directs the droplet motion along the optimal path - along the Bohmian trajectory.

 We have shown that (a) the Navier-Stokes equation together with the continuity equation can be reduced to the Schr\"{o}dinger equation assuming an incompressible irrotational fluid; (b) the Feynman path integral underlying the Schr{\"o}dinger equation can be applied for finding droplet's paths through a grating containing slits; (c) interference patterns arising behind the grating were calculated for two wavelengths given in~\cite{CouderForte:2006} - $\lambda = 4.75$~mm and $\lambda = 6.95$~mm.

 Experiments with droplets bouncing on surface of a fluid that undergoes the Faraday oscillations~\cite{CouderForte:2006, EddiEtAll:2011} can shed the light on the subtle behavior of vacuum as particles pass through.

\begin{acknowledgments}

 The author is gratefully acknowledged to O.~A.~Bykovsky and Dr.~S.~M.~Bezrukov
 for interesting discussions and series of useful remarks. 

\end{acknowledgments}



\end{document}